%% file: main_arxiv.tex
\documentclass[journal,12pt,onecolumn,draftclsnofoot,]{IEEEtran}
\IEEEoverridecommandlockouts
\IEEEoverridecommandlockouts
\usepackage{cite}
\usepackage{cite,balance}
\usepackage{amsmath,amssymb,amsfonts,nicematrix}
\usepackage{bbm}
\usepackage{xcolor}
\usepackage{algorithm}
\usepackage{algpseudocode}
\def\BibTeX{{\rm B\kern-.05em{\sc i\kern-.025em b}\kern-.08em
    T\kern-.1667em\lower.7ex\hbox{E}\kern-.125emX}}
\usepackage{nicematrix,tikz}
\usepackage{graphicx}
\usepackage{textcomp}
\usepackage{array,booktabs,makecell}
\usepackage[font=footnotesize]{caption}
\usepackage{float}
\usepackage{tikz}
\usepackage{cite,balance}
\usetikzlibrary{shapes.geometric, arrows}
\usepackage{pgfplots}
\usepackage{caption}
\usepackage{mathtools}
\usepackage{subcaption}
\usepackage{comment}
\usepackage{enumitem}
\newcommand\numberthis{\addtocounter{equation}{1}\tag{\theequation}}
\usepackage{xcolor}
\definecolor{OliveGreen}{rgb}{0,0.6,0}
\def\BibTeX{{\rm B\keri-.05em{\sc i\keri-.025em b}\keri-.08em
    T\keri-.1667em\lower.7ex\hbox{E}\keri-.125emX}}
\usepackage{multirow}

\usepackage{amsthm}
\newtheorem{remark}{Remark}

\newtheorem{theorem}{Theorem}

\newtheorem{lemma}{Lemma}

\allowdisplaybreaks
\setlist[enumerate]{wide=0pt, leftmargin=15pt, labelwidth=15pt, align=left}
\usepackage{lipsum}
\newcommand{\defeq}{\vcentcolon=}

\usepackage{caption}
\usepackage{lipsum} 
\usepackage{changepage} 
\usetikzlibrary{spy}

\usepackage{lipsum,xcolor}
\usepackage{graphicx,midfloat,capt-of}

\definecolor{bleudefrance}{rgb}{0.19, 0.55, 0.91}
\definecolor{dollarbill}{rgb}{0.52, 0.73, 0.4}
\definecolor{fuchsiapink}{rgb}{1.0, 0.47, 1.0}
\definecolor{mikadoyellow}{rgb}{1.0, 0.77, 0.05}
\definecolor{tangerine}{rgb}{0.95, 0.52, 0.0}
\definecolor{slateblue}{rgb}{0.42, 0.35, 0.8}
\definecolor{meatbrown}{rgb}{0.9, 0.72, 0.23}
\definecolor{lavenderpink}{rgb}{0.98, 0.68, 0.82}
\definecolor{lavenderrose}{rgb}{0.98, 0.63, 0.89}
\definecolor{lavendermagenta}{rgb}{0.93, 0.51, 0.93}
\definecolor{jonquil}{rgb}{0.98, 0.85, 0.37}
\definecolor{heliotrope}{rgb}{0.87, 0.45, 1.0}
\definecolor{electriclavender}{rgb}{0.96, 0.73, 1.0}
\definecolor{spirodiscoball}{rgb}{0.06, 0.75, 0.99}
\definecolor{azure(colorwheel)}{rgb}{0.0, 0.5, 1.0}

\begin{document}

\title{Cooperative Gradient Coding for Semi-Decentralized Federated Learning \\
}

\author{\IEEEauthorblockN{Shudi Weng, Chengxi Li, Ming Xiao, Mikael Skoglund\\}
\IEEEauthorblockA{\textit{School of Electrical Engineering and Computer Science} \\
\textit{KTH Royal Institute of Technology}\\
Stockholm, Sweden \\
\{shudiw,chengxli,mingx,skoglund\}@kth.se}
}

\maketitle

\begin{abstract}
Stragglers' effects are known to degrade FL performance. In this paper, we investigate federated learning (FL) over wireless networks in the presence of communication stragglers, where the power-constrained clients collaboratively train a global model by iteratively optimizing a local objective function with their local datasets and transmitting local model updates to the central parameter server (PS) through fading channels. To tackle communication stragglers without dataset sharing or prior information about the network at PS, we propose cooperative gradient coding (CoGC) for semi-decentralized FL to enable the exact global model recovery at PS. Furthermore, we conduct a thorough theoretical analysis of the proposed approach. Namely, an outage analysis of the proposed approach is provided, followed by a convergence analysis based on the failure probability of the global model recovery at PS. Nevertheless, simulation results reveal the superiority of the proposed approach in the presence of stragglers under imbalanced data distribution.
\end{abstract}

\section{Introduction}
Federated learning (FL) is a burgeoning distributed optimization paradigm in e.g., Internet-of-Things (IoT) and cloud computing applications. FL employs a privacy-preserving framework where both data collection and model training are pushed to numerous edge devices\cite{kairouz2021advances,niknam2020federated}. These edge devices (clients) collaboratively train a global model by sharing local model updates via wireless links, thereby bypassing the necessity of raw datasets sharing and consequently greatly reducing the communication overhead \cite{mammen2021federated,yin2021comprehensive}. However, limited by communication resources such as bandwidth and power constraints,  clients may suffer link disruption (due to e.g., fading and interference) and fail to upload their local model updates to the central parameter server (PS), known as \textit{communication stragglers}. 

Communication stragglers can severely degrade FL performance if not properly handled as the data distribution on a subset of clients may not represent the overall population. To this end, authors in \cite{wang2021quantized} provide a rigorous convergence analysis that quantitatively illustrates the impact of communication outages on FL convergence, revealing that the imbalanced partial participation induced by communication stragglers leads to strict sub-optimality.
To mitigate communication stragglers, \cite{saha2022colrel} proposes a semi-decentralized network that enables collaboration among clients such that the updates from the poorly connected clients can be conveyed to PS with the help of their neighbors.
However, to ensure an unbiased recovery of the global model at the PS, precise prior knowledge about the connectivity of the entire intermittent network is essential to compute the collaboration weight, which significantly increases the complexity of implementation in practical scenarios.

Analogously, distributed learning (DL) also suffers from the stragglers' effect. To improve straggler resilience, a gradient coding (GC) approach is proposed by assigning each client $s$ redundant datasets replicated from its $s$ neighbors. 
Since FL prohibits raw dataset sharing, \cite{schlegel2023codedpaddedfl} extends GC to FL by replicating coded datasets such that clients cannot discern the raw datasets of each other due to artificial randomness. However, for large datasets and a massive number of clients, the transmission of coded datasets takes excessive power and time.  To overcome the limitations of the existing methods, 
this work proposes a cooperative network based on GC mechanism to improve straggler resilience in FL systems.

\begin{figure}
    \centering
    \includegraphics[width=0.7\linewidth]{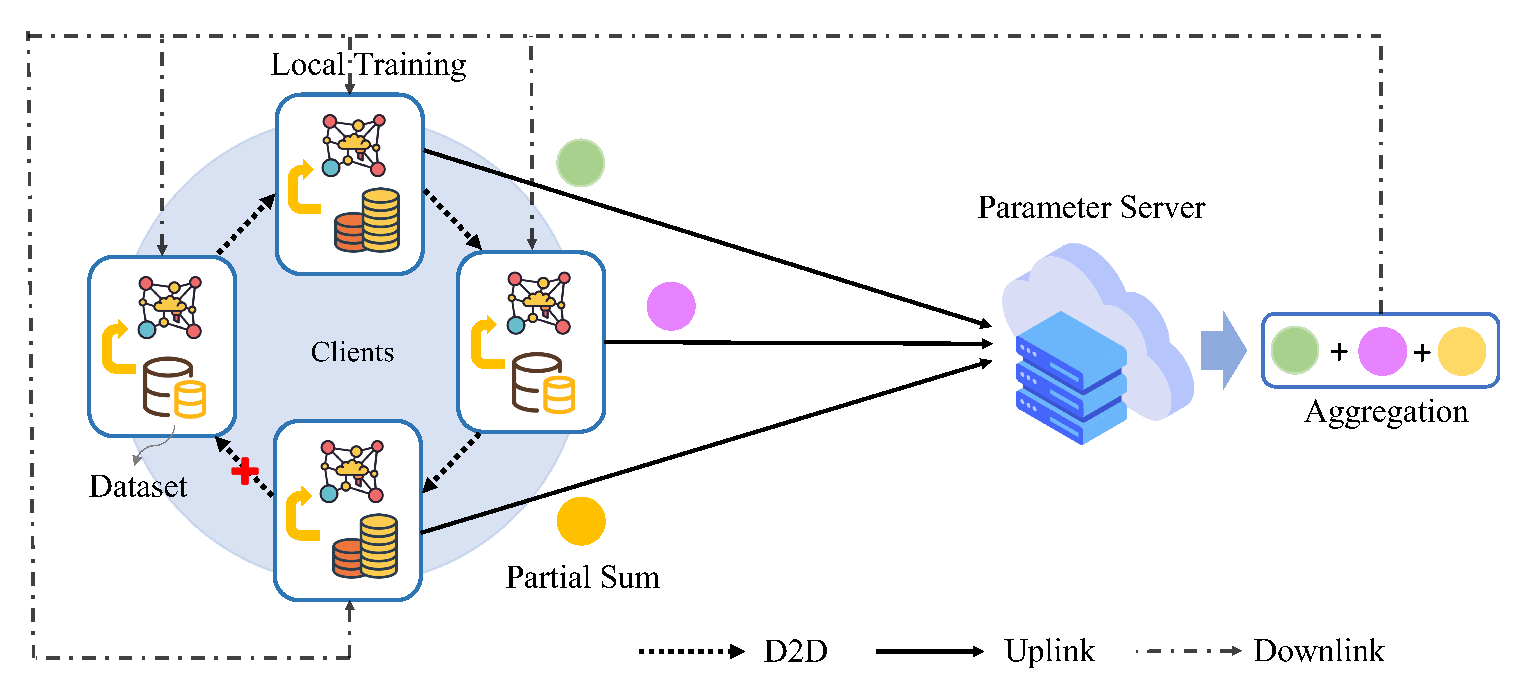}
    \vspace{-0mm}
    \caption{An illustration of the proposed approach in one training round, where the clients compute the partial sums by aggregating the local models received during the device-to-device (D2D) stage, and then transmit it to PS. PS aggregates the partial sums to recover the global model and then updates the global model with all clients.}
    \label{fig:senario}
    \vspace{-5mm}
\end{figure}



Our main contributions are summarized as follows.
\begin{itemize}
    \item We propose a cooperative gradient coding (CoGC) scheme to recover the exact global model, in which the global model either can be recovered perfectly or cannot be recovered at all each round, eliminating the need for any prior connectivity information at the PS and avoiding any type of dataset sharing. 
    \item We conduct a thorough theoretical analysis of the proposed approach over the intermittent wireless network. Namely, an outage analysis of the proposed approach is provided, followed by convergence analysis with the failure probability of the proposed approach each round.
    \item We demonstrate the effectiveness of the proposed approach through simulations. The results show that our proposed method attains better performance than baseline methods in the presence of stragglers, especially in non-i.i.d. (independent and identically distributed) settings. 
\end{itemize}
\section{Preliminaries and System Model}
\subsection{Federated Learning (FL)}\label{Sec:FedAvg}
Let $\mathcal{L}(\boldsymbol{\theta}, \xi)$ represent the loss evaluated by the learning model $\boldsymbol{\theta}\in\mathbb{R}^d$ on a data sample $\xi$. 
Consider an FL system consisting of a PS and a set of clients $\{1,\cdots,M\}$, denoted by $[M]$. 
The FL system aims to solve the following empirical risk minimization (ERM) problem: 
\begin{align*}
   \min_{\boldsymbol{\theta}\in \mathbb{R}^d} \left[F(\boldsymbol{\theta})\defeq\sum_{m=1}^{M} p_m F_m(\boldsymbol{\theta})\right],
   \label{Eq: distributed_optimization}
   \numberthis
\end{align*}
where $F(\boldsymbol{\theta})$ is defined as the global objective function, $F_m(\boldsymbol{\theta})=\frac{1}{n_m}\sum_{\xi\in \mathcal{D}_m}\mathcal{L}(\boldsymbol{\theta}, \xi)$ is the local objective function, $\mathcal{D}_m$ is the local dataset on client $m$, and $p_m=n_m/n$ denotes the learning weight with $n_m= \lvert \mathcal{D}_m \rvert$ and $n=\sum_{m=1}^M \lvert \mathcal{D}_m \rvert$.

The most popular optimization algorithm to solve (\ref{Eq: distributed_optimization}) is FedAvg \cite{mcmahan2017communication}. At $r$-th round, the following steps are executed.\\
\textbf{Broadcasting:} In the beginning, PS broadcasts the latest global model $\boldsymbol{\theta}_{r-1}$ to all clients.\\
\textbf{Training:} Client $m$ sets $\boldsymbol{\theta}_{m,r}^{0}=\boldsymbol{\theta}_{r-1}$, and performs consecutive $I$-step stochastic gradient descent (SGD). At each iteration, the local model is updated as
\begin{align*}
\boldsymbol{\theta}_{m,r}^{i}\leftarrow\boldsymbol{\theta}_{m,r}^{i-1}-\eta  \nabla F_m(\boldsymbol{\theta}_{m,r}^{i-1},\xi_{m,r}^{i}),
\numberthis
\label{eq:local_update}
\end{align*}
where $\eta>0$ is the learning rate, and $\nabla F_m(\boldsymbol{\theta}_{m,r}^{i-1},\xi_{m,r}^{i})$ is the stochastic gradient computed on data sample $\xi_{m,r}^{i}$ randomly selected at $i$-th iretation of $r$-th round on client $m$ . \\
\textbf{Transmission:} The goal of the transmission process is to convey each local model updates $\Delta\boldsymbol{\theta}_{m,r}^{I}$s to PS, where
\begin{align*}
\Delta\boldsymbol{\theta}_{m,r}^{I}=\boldsymbol{\theta}_{m,r}^{I}-\boldsymbol{\theta}_{r-1}.
\numberthis
\label{eq:local_model_update}
\end{align*}
\textbf{Aggregation:} For full participation of clients, PS performs aggregation 
and updates the global model of $r$-th round as
\begin{align*}
{\boldsymbol{\theta}}_{r}\leftarrow \boldsymbol{\theta}_{r-1} + \sum_{m=1}^{M} p_m \Delta\boldsymbol{\theta}_{m,r}^{I}.
    \numberthis
    \label{Eq:global update}
\end{align*} 

\subsection{Transmission over Wireless Networks}\label{Sec:outage}
\subsubsection{Quantized Transmission}\label{sec:SQ}
Before transmission, device $m$ needs to quantize $\Delta\boldsymbol{\theta}_{m,r}^{I}\in \mathbb{R}^d$ such that a finite number of symbols can represent the source (due to power constraint). The most commonly used technique for this purpose in FL is stochastic quantization (SQ) \cite{wang2021quantized}, whose characteristic function is given in (\ref{eq:SQ}). Assume that $\Delta\theta\in \Delta\boldsymbol{\theta}_{m,r}^{I}$ is bounded,
and let $\{c_0, c_1, \cdots, c_{2^{B}-1}\}$ be knobs uniformly distributed within $\left[\underline{\Delta\theta}, \overline{\Delta\theta}\right]$, where $\overline{(\cdot)}$ and $ \underline{(\cdot)}$ represent the upperbound and lowerbound respectively. For $\Delta\theta\in \Delta\boldsymbol{\theta}_{m,r}^{I}$, whose absolute value $\lvert \Delta\theta\rvert$ falls in $[c_l, c_{l+1})$, it is quantized by 
\begin{align*}
    \mathcal{Q}(\Delta\theta)=\begin{cases}
        \mathrm{sign}(\Delta\theta)\cdot c_l,\;\;\;\;\mathrm{w.p.}\frac{c_{l+1}-\lvert \Delta\theta\rvert}{c_{l+1}-c_l}\\
        \mathrm{sign}(\Delta\theta)\cdot c_{l+1},\;\mathrm{w.p.}\frac{\lvert \Delta\theta\rvert-c_l}{c_{l+1}-c_l},
    \end{cases}
    \numberthis
    \label{eq:SQ}
\end{align*}
where $c_l= \underline{\Delta\theta}+l\times \frac{\overline{\Delta\theta}-\underline{\Delta\theta}}{2^B-1}$
with $l=0, \cdots, 2^{B}-1$, and $B$ is the number of quantization bits. Additionally, $1$ bit is needed to indicate $\mathrm{sign}(\Delta\theta)$. 
\begin{lemma}
By adopting SQ, it holds that \cite{wang2021quantized}
\begin{align*}
&\hspace{-2mm}\mathbb{E}\left[\mathcal{Q}\left(\Delta\boldsymbol{\theta}_{m,r}\right)\right]=\Delta\boldsymbol{\theta}_{m,r} \numberthis \\
&\hspace{-2mm}\mathbb{E}\left[\lVert\mathcal{Q}\left(\Delta\boldsymbol{\theta}_{m,r}\right)-\Delta\boldsymbol{\theta}_{m,r}\rVert^2\right]\leq 
 \frac{\eta^2\delta_{m,r}^2}{(2^{B}-1)^2}\triangleq \eta^2 J_{m,r}^2
\numberthis
\end{align*}
where $\delta_{m,r}\triangleq \sqrt{\frac{1}{4} \sum_{j=1}^{d}(\overline{\nabla F_{m,r}}-\underline{\nabla F_{m,r}})^2 }$ by defining $\nabla F_{m,r}=\sum_{i=1}^{I} \nabla F_m(\boldsymbol{\theta}_{m,r}^{i-1},\xi_{m,r}^{i})$. 
\end{lemma}

\subsubsection{Semi-Decentralized Network Model}
The semi-decentralized network involves the following two stages.
\begin{enumerate}[label=(\alph*)]
    \item \hspace{-2mm}\textit{Communication between clients:} The connectivity of the links among clients can be captured by the random binary matrix $\boldsymbol{\mathcal{T}}(r)\in\{0, 1\}^{M\times M}$
    whose $(m, k)$-th entry $\tau_{mk}(r)\sim \mathrm{Bernoulli}(1-q_{mk})$, where $q_{mk}$ is the outage probability of the link from client $m$ to client $k$ and $q_{mm}=0$ since there is no transmission. 
    \item \hspace{-2mm}\textit{Communication between clients and PS:} The connectivity of the links from clients to PS can be captured by binary random vector $\boldsymbol{\tau}(r)\in\{0, 1\}^{M\times 1}$,
    whose $m$-th entry $\tau_{m}(r)\sim \mathrm{Bernoulli}(1-q_{m})$, where $q_{m}$ is the outage probability of the link from client $m$ to PS.
\end{enumerate}
\subsubsection{Outage Model of An Individual Link}
Assume orthogonal links in the semi-decentralized network, consider single transmission with the signal-to-noise ratio (SNR) $\gamma_x$ via an individual link $x$, and assume Rayleigh fading, i.e., $ h_x \sim \mathcal{CN}(0,\sigma^2_x)$, outage occurs when the channel capacity $C_x$ is less than the transmission rate $R_x$, i.e., when $C_x<R_x$, where $C_x=\frac{1}{2}\log(1+\lvert h_x\rvert^2\gamma_x)$. Or equivalently, when $\lvert h_x\rvert^2<g_x$, where $g_x=\frac{2^{2R_x}-1}{\gamma_x}$. The outage probability $q_x$ is given by 
\begin{align*}
    q_x=1-e^{-g_x/2\sigma^2_x}.
    \numberthis
    \label{eq: outage model}
\end{align*}
According to (\ref{eq: outage model}), $q_{mk}$ and  $q_{m}$ that characterize the semi-decentralized network can be computed.

\subsection{Gradient Coding (GC) in Distributed Learning (DL)}\label{Sec: GC}
To address stragglers in DL, GC is proposed in \cite{tandon2017gradient} to assign each client 
its dataset and $s$ redundant datasets by replicating its neighbors' datasets according to the non-zero pattern of the corresponding row in the cyclic 
allocation matrix $\boldsymbol{B}_{\mathrm{cyc}}\in {\mathbb{R}^{M \times M}}$ with $s+1$ non-zero entries in each row. 
Additionally, a combination matrix $\boldsymbol{A} \in {\mathbb{R}^{f \times M}}$ is defined, where $f={M \choose s}$. The patterns of non-zero entries in rows of $\boldsymbol{A}$ encompass all possible straggler patterns. The pair of $\boldsymbol{B}_{\mathrm{cyc}}$ and $\boldsymbol{A}$ is intricately designed such that 
\begin{align*}
\boldsymbol{A}\boldsymbol{B}_{\mathrm{cyc}}=\mathbf{1}_{f\times M}.
\numberthis
\label{eq: GC_property}
\end{align*}
The algorithm to generate $\boldsymbol{A}$ and $\boldsymbol{B}_{\mathrm{cyc}}$ is given in \cite{tandon2017gradient}.

After computing gradients of all allocated datasets in each round, each non-straggler device without computation error computes the weighted sum of these gradients according to the corresponding row of $\boldsymbol{B}_{\mathrm{cyc}}$ and sends it to PS. Then PS detects the straggler pattern $\boldsymbol{a}_{f_r}$ of $r$-th round where $f_r\in [f]$, i.e., the $f_r$-th row $\boldsymbol{a}_{f_r}$ of $\boldsymbol{A}$ wherein the positions of $0$s cover the indices of stragglers, mathematically, $\boldsymbol{a}_{f_r}$ should satisfy
\begin{align*}
\boldsymbol{a}_{f_r}=\boldsymbol{a}_{f_r}\bullet \boldsymbol{\tau}(r)^\top,
    \numberthis
    \label{eq:detect straggler patterns}
\end{align*}
where $\bullet$ denotes Hadamard product (element-wise).
Subsequently, PS computes the global model by combining the received weighted sum according to $\boldsymbol{a}_{f_r}$. The GC scheme is robust to any $s$ stragglers in DL, however, it is not directly applicable to FL due to the prohibition of raw dataset sharing.

\section{Cooperative Gradient Coding (CoGC)}\label{Sec:prop}

In this section, we introduce the proposed CoGC scheme. As depicted in Fig. \ref{fig:senario}, the considered FL system comprises $M$ clients and a relatively distant PS. W.L.O.G., we assume all clients are equipped with the same stochastic quantizer, the same encoder $\mathcal{E}: \mathbb{R}^d\rightarrow \mathbb{F}_p^N$ and decoder $\mathcal{E}^{-1}: \mathbb{F}_p^N \rightarrow \mathbb{R}^d $ such that they can decode each other's message according to the default systematic Gaussian codebook (for ease of outage model of a single link \cite{5595117}), where $p$ is the size of finite field, and $N$ is the block length. Nevertheless, we assume that the quantization boundary is known to all clients and PS and only $B+1$ bits representing the corresponding quantized knob and the sign need to be transmitted. Additionally, we assume the same rate $R=\frac{(B+1)d}{N}$ for all transmissions and i.i.d. Rayleigh fading for links among clients, i.e., $h_{km}\sim \mathcal{CN}(0,\sigma^2_a)$, and for links between clients and PS, i.e., $h_{m}\sim \mathcal{CN}(0,\sigma^2_b)$, respectively. The downlink channels are assumed to be error-free. 

Setting the number of preliminary communication rounds as $T$, here, we describe the learning process of the proposed CoGC at the $r$-th training round. \vspace{1mm}\\
\textbf{Broadcasting:} Unlike standard FedAvg, our design determines whether to broadcast the global model at the beginning of the $r$-th round based on whether PS successfully updated the global model in the previous $r-1$-th round. If updated, PS broadcasts the latest global model $\boldsymbol{\theta}_{r-1}$ to all clients. Otherwise, PS skips the broadcasting.
\vspace{1mm}\\ 
\textbf{Training:} If PS broadcasts $\boldsymbol{\theta}_{r-1}$, the clients set $\boldsymbol{\theta}_{m,r}^{0}=\boldsymbol{\theta}_{r-1}$ and perform $I$-step local training in (\ref{eq:local_update}). Otherwise, the clients initialize itself with its newest local model by setting $\boldsymbol{\theta}_{m,r}^{0}=\boldsymbol{\theta}_{m,r-1}^I$ and proceed with local training.\vspace{1mm}\\
\textbf{Transmission:} Prior to transmission, client $m$ computes its local model update $\Delta\boldsymbol{\theta}_{m,r}^{I}$ as in (\ref{eq:local_model_update}), and quantizes it to $\mathcal{Q} (\Delta\boldsymbol{\theta}_{m,r}^{I})$ as described in \ref{sec:SQ}. Subsequently, the following two communication stages are carried out.
\subsubsection*{(a) Device-to-Device (D2D) Communication}\label{Sec:communication1}
First, client $m$ encodes its quantized local model update $\mathcal{Q}(\Delta\boldsymbol{\theta}_{m,r}^{I})$ to message $U_{m,r}$, i.e.,  
\begin{align*}
    U_{m,r}=\mathcal{E}\left(\mathcal{Q}(\Delta\boldsymbol{\theta}_{m,r}^{I})\right).
        \numberthis
    \label{eq:encoder}
\end{align*}
Then client $m$ transmits $U_{m,r}$ at SNR $\gamma_a$ through orthogonal links, and listens from its $s$ neighbors in $\mathcal{K}_m(r)$ , where $\mathcal{K}_m(r)=\{k\vert k\neq m: b_{mk}\neq 0 \}$ and $b_{mk}$ denote the $(m,k)$-th entry in $\boldsymbol{B}_{\mathrm{cyc}}$. For each link, the outage probability is $q_a=1-e^{-g_a/2\sigma^2_a}$ as illustrated in (\ref{eq: outage model}). The binary connectivity matrix $\boldsymbol{\mathcal{T}}_{\mathrm{cyc}}(r)$ characterizing this D2D network of ring topology has the same cyclic patterns as $\boldsymbol{B}_{\mathrm{cyc}}$, i.e., the entries $\tau_{mk}$s in $\boldsymbol{\mathcal{T}}_{\mathrm{cyc}}(r)$ is of the form
\begin{align*}
    \tau_{mk}=\begin{cases}
        \mathrm{Bernoulli}(1-q_a) \;\;\;\;\;\; \mathrm{if} \; b_{mk}\neq 0\\
        0 \;\;\;\;\;\;  \;\;\;\;\;\;  \;\;\;\;\;\;  \;\;\;\;\;\;  \;\;\;\;\;\;  \;
        \mathrm{if}\; b_{mk}=0
    \end{cases}.
    \numberthis
    \label{eq:D2D_connectivity}
\end{align*}
Consequently, client $m$ can decode
\begin{align*}
    \mathcal{Q}(\Delta\boldsymbol{\theta}_{m,r}^{I})=\mathcal{E}^{-1}\left(U_{m,r}\right)
    \numberthis
    \label{eq: decoder}
\end{align*}
from its neighbors in $\Tilde{\mathcal{K}}_m(r)$ with good connectivity, where $\Tilde{\mathcal{K}}_m(r)=\{k\vert k\neq m: \tau_{mk}\neq 0 \}$. 

If client $m$ successfully decodes all messages from its $s$ neighbors, it computes the partial sum of the decoded updates according to $\boldsymbol{B}_{\mathrm{cyc}}$. Mathematically, for client $m\in \Breve{\mathcal{K}}(r)$, where $\Breve{\mathcal{K}}(r)=\{m\vert \Tilde{\mathcal{K}}_m(r)=\mathcal{K}_m(r) \}$, it computes
\begin{align*}
    \boldsymbol{s}_{m,r}=\sum_{k=1}^M b_{mk}p_{k}\mathcal{Q}(\Delta\boldsymbol{\theta}_{k,r}^I).
    \label{eq:partial sum}
    \numberthis
\end{align*}
Otherwise, client $m$ does not compute anything and stays silent during the later steps of the $r$-th round training. 

It is worth noting that the distant PS requires significantly higher power than clients to achieve reliable communication, and thus cannot decode messages at this stage.


\subsubsection*{(b) Device-to-PS (D2P) Communication}\label{Sec:communication2} 
For client $m\in \Breve{\mathcal{K}}(r)$, it encodes the partial sum $\boldsymbol{s}_{m,r}$ to message $V_{m,r}$, i.e.,
\begin{align*}
V_{m,r}=\mathcal{E}\left(\boldsymbol{s}_{m,r}\right).
   \numberthis
\end{align*}
Subsequently, it transmits $V_{m,r}$ to PS at SNR $\gamma_b$ through orthogonal links. The binary connectivity vector $\boldsymbol{\tau}(r)$ characterizing these D2P links can be modeled as i.i.d. Bernoulli r.v.s, i.e., $\tau_m(r)\sim \mathrm{Bernoulli}(1-q_b)$, where $q_b$ follows (\ref{eq: outage model}). For $m\in \hat{\mathcal{K}}(r)$, where $\hat{\mathcal{K}}(r)=\{m\vert \tau_m(r)=1\}$, PS can decode
\begin{align*}
\boldsymbol{s}_{m,r}=\mathcal{E}^{-1}\left(V_{m,r}\right).
    \numberthis
\end{align*}
\textbf{Aggregation:}
If $\lvert\hat{\mathcal{K}}(r)\rvert\geq M-s$, PS can successfully compute the global model update $\Delta{\boldsymbol{\theta}}_r$ by combing the received partial sums $\boldsymbol{s}_{m,r}$s, according to the detected straggler pattern $\boldsymbol{a}_{f_r}$ in (\ref{eq:detect straggler patterns}). The global model update $\Delta{\boldsymbol{\theta}}_r$ is computed as 
\begin{align*}
    \Delta{\boldsymbol{\theta}}_r=\sum_{m=1}^M a_{f_r,m} \boldsymbol{s}_{m,r},
    \label{eq: global_update_at_PS}
    \numberthis
\end{align*}
where $a_{f_r,m}$ is the $m$-th element in $\boldsymbol{a}_{f_r}$. By the feature of GC scheme in (\ref{eq: GC_property}), it can easily shown that (\ref{eq: global_update_at_PS}) is exactly $\sum_{m=1}^M p_m \mathcal{Q}(\Delta\boldsymbol{\theta}_{m,r}^{I})$. After commuting $\Delta{\boldsymbol{\theta}}_r$, PS updates the global model as in (\ref{Eq:global update}). If $\lvert\hat{\mathcal{K}}(r)\rvert<M-s$, PS cannot update anything, without possibly receiving a distorted global model caused by partial participation. If the number of total stragglers exceeds $s$, the global model is not updated at the $r$-th round.

The training stops when the following two conditions are satisfied:  $(i)$ the total number of executed training rounds reaches $T$, and $(ii)$ the global model of the last executed round is successfully updated. 

Let $P_O$ be the overall outage probability, which refers to the probability of global model recovery failure at PS each round. CoGC is mirrored by the learning process given in Algorithm 1, based on which we conduct the convergence analysis in \ref{Sec: convergence_analysis}. 
In the mirror learning process, the number of consecutive training rounds $R_r$ between $r-1$-th and $r$-th successful recovery follows a Geometrical distribution, that is, clients perform $R_r I$-step SGD where $R_r\sim \mathrm{Geo}(1-P_O)$. Training stops at the first time $\sum_{r=1}^{n}R_r\geq T$, and $T'=\min\{n\in \mathbb{Z}:\sum_{r=1}^{n}R_r\geq T\}$ is the executed training rounds.



\begin{algorithm}\label{algo}
\caption{An equivalent mirror FL process}\label{alg:equiv}
\begin{algorithmic}[1]
\State{\textbf{Initialize} $\boldsymbol{\theta}_0=0$, $r=1$, $R_1=1$}
\While{$\sum_{j=1}^{r}R_j\leq T$}
\State{PS broadcasts $\boldsymbol{\theta}_{r-1}$}
\State{$R_r=k$  w.p.  $P_O^{k-1}(1-P_O)$}
\For{$m=1,\cdots,M$}
\State{client $m$ performs $R_r I$-step SGD ($\boldsymbol{\theta}_{m,r}^{0}=\boldsymbol{\theta}_{r-1}$)}
\State{client $m$ send  $\mathcal{Q}(\Delta\boldsymbol{\theta}_{m,r}^{R_r I})$ to device $k$ 
}
\If{client $m$ can computes $\boldsymbol{s}_{m,r}$ in  (\ref{eq:partial sum})}
\State {client $m$ transmits $\boldsymbol{s}_{m,r}$ to PS}
\EndIf
\EndFor
\State{PS computes $\Delta\boldsymbol{\theta}_r$ in (\ref{eq: global_update_at_PS}) and updates $\boldsymbol{\theta}_r$ in (\ref{Eq:global update}})
\State{$r=r+1$}
\EndWhile
\end{algorithmic}
\end{algorithm}
\section{Performance Analysis}
\subsection{Outage Analysis}
This section focuses on analyzing the overall outage with the single transmission outage model. For simplicity, assume $\gamma_b=\frac{\sigma_a^2}{\sigma_b^2}\gamma_a$ such that $q_a=q_b=q$. To this end, let us consider the following three sub-cases of the overall outage.

\subsubsection{No Straggler in D2D Stage} Assume no straggler occurs in D2D stage, i.e., every device successfully recovers $s$ updates of its neighbors. This event occurs w.p. $\left(\left(1-q\right)^s\right)^M$. The overall outage happens when $v_1 \geq s+1$ device-to-PS links are down, so the overall outage probability is
\begin{align*}
    P_1=\left(1-q\right)^{sM}\sum_{v=s+1}^{M} {M\choose v_1} q^{v_1}(1-q)^{M-v_1}.
    \numberthis
\end{align*}
\subsubsection{$v_1 \geq s+1$ Stragglers in D2D Stage} If there are $v_1 \geq s+1$ stragglers in D2D stage, 
the overall outage happens for sure regardless of the device-to-PS link condition. A device in D2D stage is a straggler if at least $1$ link between the device and its neighbors is down. The probability of a device being a straggler is $1-(1-q)^s$. The overall outage probability is 
\begin{align*}
\hspace{-2mm}P_2=\hspace{-1mm}\sum_{v=s+1}^{M} \hspace{-1mm} {M \choose v} \left(1-(1-q)^s \right)^{v_1}\left( (1-q)^s\right)^{M-v_1}.
\numberthis
\end{align*}
\subsubsection{1$\sim$s Stragglers in D2D Stage }
Suppose that there are $v_1\in [s]$ stragglers in D2D stage, the GC scheme is dysfunctional if there are $v_2 \geq s-v_1+1$ stragglers among $M-v_1$ clients in D2P stage. Thus, the overall outage probability is
\begin{align*}
    P_{3}&=\sum_{v_1=1}^{s}{M\choose v_1}\left(1-(1-q)^s \right)^{v_1}\left( (1-q)^s\right)^{M-v_1}\sum_{v_2=s-v_1+1}^{M-v_1} {M-v_1 \choose v_2} q^{v_2}(1-q)^{M-v_1-v_2}.
    \numberthis
\end{align*}
Since sub-cases $\textit{1)}, \textit{2)} ,\textit{3)}$ are non-overlapping, $P_{O}$ is given by
\begin{align*}
    P_{O}=P_1+P_2+P_3.
    \numberthis
\end{align*}
\subsection{Non-Convex Convergence Rate Analysis}\label{Sec: convergence_analysis}
To start with, we outline the assumptions used in the convergence analysis \cite{lian2017can}\cite{9515709}\cite{wang2020tackling}. 
\begin{itemize}
    \item[A.1] 
    Each local objective function is bounded by $F_m(x)\geq F^\star$ and is differentiable, its gradient $\nabla F_m(x)$ is L-smooth, i.e., $\lVert \nabla F_m(x)-\nabla F_m(y) \rVert\leq L\lVert x-y \rVert$, $\forall i \in [M]$.
    \item[A.2] 
    The local stochastic gradient is an unbiased estimation, i.e., $\mathbb{E}_\xi[\nabla F_m(x,\xi)]=\nabla F_m(x)$, and has bounded data variance $\mathbb{E}_\xi[\lVert \nabla F_m(x,\xi)-\nabla F_m(x)\rVert^2]\leq \sigma^2 $, $\forall i \in [M]$. 
    \item[A.3] 
    The dissimilarity between $\nabla F_m(x)$ and $\nabla F(x)$ is bounded, i.e., $\mathbb{E}[\lVert \nabla F_m(x)-\nabla F(x)\rVert^2]\leq D_m^2$, $\forall i \in [M]$. 
\end{itemize}

\begin{theorem}
Under A.1$\sim$A.3, for a given number of $I$ iterations on device $m\in [M]$, by adopting the proposed approach in which the probability of unsuccessful recovery of the global model is $P_O$, if the preliminary training rounds $T$ is chosen such that the learning rate $\eta=\frac{1}{L}\sqrt{\frac{M}{T}}$ is small enough, it yields that the optimality gap converges w.p. $p\rightarrow 1$.
\begin{align*}
    &\min_{r\in [T']}\mathbb{E}\left[\left\lVert\nabla F(\boldsymbol{\theta}_{r}^{0}) \right\rVert^2\right]\leq 2(1-P_O)\cdot\\
    &\hspace{9mm}\Bigg\{\mathcal{O}\left(\frac{1}{1-P_O}\frac{L\left(F^\star-F(\boldsymbol{\theta}_{r}^0)\right)}{\sqrt{MT}I}\right)\\
&\hspace{5mm}+\mathcal{O}\left(\left(\frac{2I(1+P_O)}{(1-P_O)^2}\sqrt{\frac{M}{T}}+\frac{MC_1}{L^2T}\right)\sum_{m=1}^M p_m D_m^2\right)\\
&\hspace{5mm}+\mathcal{O}\left(\left(\frac{\sum_{m=1}^M p_m^2}{2(1-P_O)}\sqrt{\frac{M}{T}} +\frac{MC_2}{L^2T}\right)\sigma^2\right)\\
    &\hspace{5mm}+\mathcal{O}\left(\frac{1}{2I}\sqrt{\frac{M}{T}}\frac{\sum_{r\in [(1-P_O)T]}\sum_{m=1}^M p_m^2 J_{m,r}^2}{(1-P_O)T}\right)\Bigg\}
    \numberthis
\end{align*}
\end{theorem}
\begin{remark}
Theorem 1 can be easily extended to patch SGD with $b$ nodes by replacing $\sigma^2$ by $\sigma^2/b$.
\end{remark}

\section{Simulation}
\begin{figure*}[!htb]
 \centering
     \scalebox{.6}{\input{outage} }
    \caption{Outage probability of an individual link and overall outage probability in terms of SNR.} 
    \label{fig:outage}
\end{figure*}

\begin{figure*}[!htb]
  \centering
    \scalebox{.6}{ \input{test_acc_iid} }
    \caption{Test accuracy comparison of four methods under different SNRs 
    in i.i.d. setting. }
    \label{fig:test_acc_iid}
\end{figure*}

\begin{figure*}[!htb]
    \centering
   \scalebox{.6}{ \input{test_acc_noniid_new} }
    \caption{Test accuracy comparison of four methods under different data imbalances 
    in non-i.i.d. setting. } 
     \label{fig:test_acc_noniid_new}
\end{figure*}

In the simulation, the number of clients is set to $M=10$. Each device employs SQ with $B=8$ quantization bits and boundary values of SQ remain the same each round.
We validate CoGC on the MNIST dataset, distributing an equal amount of data to each device, and evaluating the performance of the following four methods under different settings choosing $\sigma_a=1$, $\sigma_b=0.2$, $\gamma_b=\frac{\sigma_a^2}{\sigma_b^2}\gamma_a$, i.e., $q_a=q_b=q$.
\begin{enumerate}[label=(\roman*)]
\item Quantized FL (QFL) employing digital/analog transmission with perfect links\cite{9515709}, i.e., when $\gamma_a=\infty$. 
\item Our proposed method employing digital transmission, i.e., CoGC 
under $\gamma_a=3/5$ per device and $R=0.2$. 
\item Non-blind FL employing digital transmission \cite{wang2021quantized} with the same setting in (ii).
\item Blind FL with the same setting in (ii), which refers to the scenarios where the PS is unaware of the identity of received clients, such as amplify-and-forward. 
\end{enumerate} 

The preliminary number of training rounds is set to $T=20$, the number of local iterations $I=5$, the patch size per iteration is set to $\lvert \boldsymbol\xi_{m,r}^{i} \rvert=1024$ and the learning rate is set to $\eta=0.01$. The classifier model is implemented using a 4-layer convolutional neural network (CNN) with SGD optimizer that consists of two convolution layers with 10 and 20 output channels respectively followed by 2 fully connected layers. The dimension of learning parameter $d>10^4$. 

The outage probability $q$ of individual links and the overall outage probability $P_O$ in terms of SNR are plotted in Fig. \ref{fig:outage}. From Fig. \ref{fig:outage}, it is foreseeable that the prevalence of stragglers
in low SNR scenarios is higher compared to high SNR scenarios, attributed to an elevated outage probability. $P_O$ decreases with increased SNR and $s$, which indicates that $s$ can be set smaller with larger SNR, and vice versa. In our simulation, $s$ is set to 5 for $\gamma_a=5$, and 7 for $\gamma_a=3$.

The average test accuracy of the global model over multiple runs at each round is plotted in Fig. \ref{fig:test_acc_iid} and Fig. \ref{fig:test_acc_noniid_new} for i.i.d case and  non-i.i.d. case, respectively. Under i.i.d data distribution, the training samples are shuffled and uniformly assigned to $M=10$ clients. Under non-i.i.d. data distribution, each device is assigned 5 classes of data and 1 class of data respectively to achieve different extents of data imbalance. To ensure an equitable comparison, we truncate the outcome at $20$ rounds. From Fig. \ref{fig:test_acc_iid}, 
we observe that under i.i.d. data distribution, the performance of blind FL deteriorates with decreased SNR due to more frequent stragglers. The non-blind FL performs well in the presence of stragglers due to homogeneous data distribution across clients. However, with the increasing data imbalance, as shown in Fig. \ref{fig:test_acc_noniid_new}, the non-blind FL shows a slower convergence and larger generalization gap. The performance of blind FL degrades significantly with more severe data imbalance across clients. In both i.i.d. and non-i.i.d. cases, our proposed algorithm effectively handles stragglers and almost achieves the same performance as the ideal scenarios with infinite SNR.

\section{Conclusion}
In this paper, we proposed a cooperative network based on GC for semi-decentralized FL to recover the exact global model for each round, yielding excellent straggler handling ability without prior information about the networks.



\balance
\input{reference.bbl}

\appendices
\section{Proof of Theorem 1}
\begin{proof}
Assume that the $R_r-1$ overall outages happen before  successful aggregation, that is, clients perform 
$R_r I$-step consecutive SGD 
before the successful aggregation during the . According to A.1, it holds that
  \begin{align*}
      &\mathbb{E}\left[F(\boldsymbol{\theta}_{r+1}^0)\right]-\mathbb{E}\left[F(\boldsymbol{\theta}_{r}^0)\right]
     \overset{\text{A.1}}\leq\mathbb{E}\left[\left\langle \nabla F(\boldsymbol{\theta}_{r}^0), \boldsymbol{\theta}_{r+1}^0-\boldsymbol{\theta}_{r}^0 \right\rangle\right]
      +\frac{L}{2}\mathbb{E}\left[\lVert \boldsymbol{\theta}_{r+1}^0-\boldsymbol{\theta}_{r}^0 \rVert^2\right]. \label{eq:theorem 1_uniform_tau}
      \numberthis
  \end{align*}
Next, we present several useful lemmas to prove Theorem 1.

\begin{lemma}
Under A.1$\sim$A.3, it holds that
    \begin{align*}
        &\mathbb{E}\left[\left\langle \nabla F(\boldsymbol{\theta}_{r}^0), \boldsymbol{\theta}_{r+1}^0-\boldsymbol{\theta}_{r}^0 \right\rangle\right]
        \leq-\frac{1}{2}\eta R_r I\mathbb{E}\left[\left\lVert \nabla F(\boldsymbol{\theta}_{r}^0) \right\rVert^2\right]
        +\frac{1}{2}\eta L^2\sum_{i=0}^{R_r I-1}\sum_{m=1}^M p_m  \mathbb{E}\left[\left\lVert \boldsymbol{\theta}_{m,r}^0 - \boldsymbol{\theta}_{m,r}^{i} \right\rVert^2\right].
        \numberthis
    \end{align*}
\end{lemma}
\begin{proof}
    Proof of L.2 is provided in Appendix B.
\end{proof}
\begin{lemma}
Under A.1$\sim$A.3, it holds that
    \begin{align*}
      &\mathbb{E}\left[\lVert \boldsymbol{\theta}_{r+1}^0-\boldsymbol{\theta}_{r}^0 \rVert^2\right]\leq \eta^2\sum_{m=1}^M p_m^2 J_{m,r}^2+\eta^2 R_r I \sigma^2 \sum_{m=1}^M p_m^2+2\eta^2 R_r I\sum_{m=1}^M p_m \sum_{i=0}^{R_r I-1} L^2\mathbb{E}\left[\left\lVert\boldsymbol{\theta}_{m,r}^{i}-\boldsymbol{\theta}_{r}^{0}\right\rVert^2\right]\\
      &\hspace{3.5cm}
      +4\eta^2 R_r^2 I^2\hspace{-1mm}\sum_{m=1}^M p_m D_m^2+4\eta^2 R_r^2 I^2 \mathbb{E}\left[\left\lVert\nabla F(\boldsymbol{\theta}_{r}^{0}) \right\rVert^2\right].
      \numberthis
    \end{align*}
\end{lemma}
\begin{proof}
    Proof of L.3 is provided in Appendix C.
\end{proof}
\begin{lemma}
Under A.1$\sim$A.3, it holds that
    \begin{align*}
    &\sum_{i=0}^{R_r I-1}  \mathbb{E}\left[\left\lVert \boldsymbol{\theta}_{m,r}^0- \boldsymbol{\theta}_{m,r}^{i} \right\rVert^2\right]
    \leq\frac{\frac{1}{2}R_r I(R_r I+1)}{1-R_r I(R_r I+1)\eta^2L^2}\eta^2 \sigma^2
    +\frac{\frac{2}{3}R_r I(R_r I+1)(2R_r I+1)}{1-R_r I(R_r I+1)\eta^2L^2} \eta^2 D_m^2\\
    &\hspace{5cm}+\frac{\frac{2}{3}R_r I(R_r I+1)(2R_r I+1)}{1-R_r I(R_r I+1)\eta^2L^2} \eta^2 \mathbb{E}\left[\left\lVert \nabla F(\boldsymbol{\theta}_{r}^{0})\right\rVert^2\right].
    \numberthis
    \label{eq:lemma3}
    \end{align*}
\end{lemma}
\begin{proof}
    Proof of L.4 is provided in Appendix D.
\end{proof}

By substituting L.2 and L.3 into for into (\ref{eq:theorem 1_uniform_tau}), we obtain 
\begin{align*}
    &\mathbb{E}[F(\boldsymbol{\theta}_{r+1}^0)]-\mathbb{E}[F(\boldsymbol{\theta}_{r}^0)]\leq 2\eta^2 R_r^2 I^2L\sum_{m=1}^M p_m D_m^2
    +\left(2\eta^2R_r^2 I^2L -\frac{1}{2}\eta R_r I \right) \mathbb{E}\left[\left\lVert\nabla F(\boldsymbol{\theta}_{r}^{0}) \right\rVert^2\right]\\
    &+(\frac{1}{2}\eta L^2+\eta^2 L^3 R_r I)\sum_{m=1}^M p_m \sum_{i=0}^{R_r I-1} \mathbb{E}\left[\left\lVert\boldsymbol{\theta}_{m,r}^{i}-\boldsymbol{\theta}_{r}^{0}\right\rVert^2\right]
    +\frac{1}{2}\eta^2 L\sum_{m=1}^M p_m^2 J_{m,r}^2+\frac{1}{2}\eta^2 L R_r I \sigma^2 \sum_{m=1}^M p_m^2 ,
    \numberthis
\end{align*}
Utilize results from L.4, re-arrange the terms, divide both sides by $\eta I$, and average over the executed $T'$ rounds, we obtain
\begin{align*}
    &\frac{1}{T'}\sum_{r\in [T']} H_1\mathbb{E}\left[\left\lVert\nabla F(\boldsymbol{\theta}_{r}^{0}) \right\rVert^2\right]\leq
    \frac{1}{T'}H_2\left(F^\star-F(\boldsymbol{\theta}_{r}^0)\right)+\frac{1}{T'}\sum_{r\in [T']}H_3\sum_{m=1}^M p_m D_m^2\\
    &\hspace{5.3cm}+\frac{1}{T'}\sum_{r\in [T']}H_4 \sigma^2+\frac{1}{T'}\sum_{r\in [T']}H_5\sum_{m=1}^M p_m^2 J_{m,r}^2,
\numberthis
\label{eq:theorem1_tau_pre}
\end{align*}
where
\begin{align*}
    &H_1=\frac{1}{2}R_r-H_3, \hspace{3mm} H_2=\frac{1}{\eta I },\hspace{3mm} H_5=\frac{\eta L}{2I},\\
    &H_3=2\eta R_r^2 I L+\eta^2\underbrace{\frac{\frac{2}{3}L^2R_r(R_r I+1)(2R_r I+1)(\frac{1}{2}+\eta R_r I L)}{1-R_r I(R_r I+1)\eta^2L^2}}_{c_1},\\
    &H_4=\frac{1}{2}\eta R_r L \sum_{m=1}^M p_m^2+\eta^2\underbrace{\frac{\frac{1}{2}L^2R_r(R_r I+1)(\frac{1}{2}+\eta R_r I L)}{1-R_r I(R_r I+1)\eta^2L^2}}_{c_2}.
\end{align*}
 By ratio test in (\ref{ratio test}), we verify $\mathbb{E}[c_1]$ converges to finite $C_1$.
\begin{align*}
    &\mathbb{E}[c_1]=\sum_{R_r=1}^{\infty}\underbrace{c_1(1-P_O)P_O^{R_r-1}}_{c_3(R_r)}\numberthis\\
    &\lim_{R_r\rightarrow \infty}\frac{c_3(R_r+1)}{c_3(R_r)}=P_O<1
    \numberthis
    \label{ratio test}
\end{align*}
Similarly, we can verify $\mathbb{E}[c_2]$ converges to finite $C_2$.

Since $R_r$ follows geometrical distribution, it can be shown that $\mathbb{E}[R_r]=\frac{1}{1-P_O}$ and $\mathbb{E}[R_r^2]=\frac{1+P_O}{(1-P_O)^2}$.
Subsequently, 
\begin{align*}
    &\mathbb{E}[H_3]=2\eta I L\frac{1+P_O}{(1-P_O)^2}+\eta^2C_1, \numberthis\\
    &\mathbb{E}[H_4]=\frac{1}{2}\eta L \sum_{m=1}^M p_m^2\frac{1}{1-P_O}+\eta^2C_2,\numberthis\\
    &\mathbb{E}[H_1]=\frac{1}{2(1-P_O)}-\mathbb{E}[H_3]\numberthis.
\end{align*}

When $T\rightarrow\infty$, $R_r$ are i.i.d. geometrical variables, By L.2 and law of large numbers $T'=\left \lceil\frac{T}{\mathbb{E}[R_r]}\right \rceil \approx (1-P_O)T$ w.p. $p\rightarrow 1$. When $P_O\ll 1$
, (\ref{eq:theorem1_tau_pre}) can be approximated as
\begin{align*}
    &\mathbb{E}[H_1]\min_{r\in [(1-P_O)T]}\mathbb{E}\left[\left\lVert\nabla F(\boldsymbol{\theta}_{r}^{0}) \right\rVert^2\right]\leq
    \frac{H_2}{(1-P_O)T}\left(F^\star-F(\boldsymbol{\theta}_{r}^0)\right)+\mathbb{E}[H_3]\sum_{m=1}^M p_m D_m^2\\
    &\hspace{6.2cm}+\mathbb{E}[H_4] \sigma^2+\frac{H_5}{(1-P_O)T}\sum_{r\in [(1-P_O)T]}\sum_{m=1}^M p_m^2 J_{m,r}^2.
\numberthis
\label{eq:theorem1_tau_pre2}
\end{align*}
By choosing learning rate $\eta=\frac{1}{L}\sqrt{\frac{M}{T}}$, we complete the proof. 
\end{proof}
  \section{proof of lemma 2} \label{Appendix:Lemma2}
  The term $\mathbb{E}\left[\left\langle \nabla F(\boldsymbol{\theta}_{r}^0), \boldsymbol{\theta}_{r+1}^0-\boldsymbol{\theta}_{r}^0 \right\rangle\right]$ is bounded by
  \begin{align*}
     &\mathbb{E}\left[\left\langle \nabla F(\boldsymbol{\theta}_{r}^0), \boldsymbol{\theta}_{r+1}^0-\boldsymbol{\theta}_{r}^0 \right\rangle\right]\\
    &=\mathbb{E}\left[\left\langle \nabla F(\boldsymbol{\theta}_{r}^0), \sum_{j=1}^M a_{kj}\sum_{m=1}^M b_{jm} p_m \mathcal{Q}\left(\Delta \boldsymbol{\theta}_{r,m}^{R_r\tau-1}\right) \right\rangle\right]\\
    &\overset{\text{L.1}}=-\eta\mathbb{E}\left[\left\langle \nabla F(\boldsymbol{\theta}_{r}^0), \sum_{m=1}^M p_m \sum_{i=0}^{R_r\tau-1} \nabla F_m(\boldsymbol{\theta}_{m,r}^{i},\xi_{m,r}^{i}) \right\rangle\right]\\
     &\overset{\text{(a)}}=-\eta\sum_{i=0}^{R_r\tau-1}\mathbb{E}\left[\left\langle \nabla F(\boldsymbol{\theta}_{r}^0), \sum_{m=1}^M p_m  \nabla F_m(\boldsymbol{\theta}_{m,r}^{i}) \right\rangle\right]\\
     &\overset{\text{(b)}}=-\frac{1}{2}\eta\sum_{i=0}^{R_r\tau-1}\mathbb{E}\left[\left\lVert \nabla F(\boldsymbol{\theta}_{r}^0) \right\rVert^2\right]
     -\frac{1}{2}\eta\sum_{i=0}^{R_r\tau-1}\mathbb{E}\left[\left\lVert \sum_{m=1}^M p_m  \nabla F_m(\boldsymbol{\theta}_{m,r}^{i}) \right\rVert^2\right]\\
     &\hspace{5mm}+\frac{1}{2}\eta\sum_{i=0}^{R_r\tau-1}\mathbb{E}\left[\left\lVert \nabla F(\boldsymbol{\theta}_{r}^0)-\sum_{m=1}^M p_m  \nabla F_m(\boldsymbol{\theta}_{m,r}^{i}) \right\rVert^2\right]\\
     &\overset{\text{(c)}}\leq-\frac{1}{2}\eta R_r\tau
     \mathbb{E}\left[\left\lVert \nabla F(\boldsymbol{\theta}_{r}^0) \right\rVert^2\right]
     +\frac{1}{2}\eta\sum_{i=0}^{R_r\tau-1}\sum_{m=1}^M p_m \mathbb{E}\left[\left\lVert \nabla F_m(\boldsymbol{\theta}_{r}^0) -  \nabla F_m(\boldsymbol{\theta}_{m,r}^{i}) \right\rVert^2\right]\\
     &\overset{\text{A.1}}\leq-\frac{1}{2}\eta R_r\tau\mathbb{E}\left[\left\lVert \nabla F(\boldsymbol{\theta}_{r}^0) \right\rVert^2\right]
     +\frac{1}{2}\eta L^2\sum_{i=0}^{R_r\tau-1}\sum_{m=1}^M p_m  \mathbb{E}\left[\left\lVert \boldsymbol{\theta}_{m,r}^0 - \boldsymbol{\theta}_{m,r}^{i} \right\rVert^2\right]
    \numberthis
  \end{align*}
  where $\text{(a)}$ follows A.2 and property of inner product, $\text{(b)}$ follows the fact that $\langle a,b \rangle=\frac{1}{2}\lVert a \rVert^2+\frac{1}{2}\lVert b \rVert^2-\frac{1}{2}\lVert a-b \rVert^2$, $\text{(c)}$ follows $\nabla F(\boldsymbol{\theta}_{r}^0)=\sum_{m=1}^M p_m \nabla F_m(\boldsymbol{\theta}_{r}^0)$ and the convexity of $l_2$-norm. 
  \section{proof of lemma 3} \label{Appendix:Lemma3}
  The term $\mathbb{E}\left[\lVert \boldsymbol{\theta}_{r+1}^0-\boldsymbol{\theta}_{r}^0 \rVert^2\right]$ is bounded by
  \begin{align*}
      &\mathbb{E}\left[\lVert \boldsymbol{\theta}_{r+1}^0-\boldsymbol{\theta}_{r}^0 \rVert^2\right]
      =\mathbb{E}\left[\left\lVert \sum_{m=1}^M p_m  \mathcal{Q}\left(\Delta \boldsymbol{\theta}_{r,m}^{R_r\tau-1}\right) \right\rVert^2\right]\\
      &\overset{\text{(d)}}=\mathbb{E}\left[\left\lVert \sum_{m=1}^M p_m \left( \mathcal{Q}\left(\Delta \boldsymbol{\theta}_{r,m}^{R_r\tau-1}\right) -\Delta \boldsymbol{\theta}_{r,m}^{R_r\tau-1} \right) \right\rVert^2\right]
      +\mathbb{E}\left[\left\lVert \sum_{m=1}^M p_m\Delta \boldsymbol{\theta}_{r,m}^{R_r\tau-1} \right\rVert^2\right]\\
      &\overset{\text{(e)}}=\sum_{m=1}^M p_m^2 \mathbb{E}\Bigg[\Big\lVert \mathcal{Q}\left( \Delta \boldsymbol{\theta}_{r,m}^{R_r\tau-1} \right)-\Delta \boldsymbol{\theta}_{r,m}^{R_r\tau-1} \Big\rVert^2\Bigg]
      +\eta^2 \mathbb{E}\left[\left\lVert \sum_{m=1}^M p_m\sum_{i=0}^{R_r\tau-1} \nabla F_m(\boldsymbol{\theta}_{m,r}^{i},\xi_{m,r}^{i}) \right\rVert^2\right]\\
      &\overset{\text{(f)}}=\sum_{m=1}^M p_m^2 J_{m,r}^2+\eta^2 \mathbb{E}\left[\left\lVert \sum_{m=1}^M p_m\sum_{i=0}^{R_r\tau-1} \nabla F_m(\boldsymbol{\theta}_{m,r}^{i}) \right\rVert^2\right]\\
      &\hspace{5mm}+\eta^2\mathbb{E}\left[\left\lVert \sum_{m=1}^M p_m\hspace{-2mm}\sum_{i=0}^{R_r\tau-1} \left(\nabla F_m(\boldsymbol{\theta}_{m,r}^{i},\xi_{m,r}^{i})-\nabla F_m(\boldsymbol{\theta}_{m,r}^{i})\right) \right\rVert^2\right]\\
      &\overset{\text{(g)}}=\sum_{m=1}^M p_m^2 J_{m,r}^2+\eta^2 \mathbb{E}\left[\left\lVert \sum_{m=1}^M p_m\sum_{i=0}^{R_r\tau-1} \nabla F_m(\boldsymbol{\theta}_{m,r}^{i}) \right\rVert^2\right]\\
      &\hspace{5mm}+\eta^2 \sum_{m=1}^M p_m^2\hspace{-2mm}\sum_{i=0}^{R_r\tau-1}  \mathbb{E}\left[\left\lVert \left(\nabla F_m(\boldsymbol{\theta}_{m,r}^{i},\xi_{m,r}^{i})-\nabla F_m(\boldsymbol{\theta}_{m,r}^{i})\right) \right\rVert^2\right]\\
      &\overset{\text{A.2}}\leq \sum_{m=1}^M p_m^2 J_{m,r}^2+\eta^2 R_r\tau \sigma^2 \sum_{m=1}^M p_m^2+\eta^2 \sum_{m=1}^M p_m
      \mathbb{E}\left[\left\lVert \sum_{i=0}^{R_r\tau-1} \nabla F_m(\boldsymbol{\theta}_{m,r}^{i})-\nabla F_m(\boldsymbol{\theta}_{m,r}^{0})+\nabla F_m(\boldsymbol{\theta}_{m,r}^{0}) \right\rVert^2\right]\\
      &\overset{\text{(h)}}\leq \sum_{m=1}^M p_m^2 J_{m,r}^2+\eta^2 R_r\tau \sigma^2 \sum_{m=1}^M p_m^2 
      +2\eta^2 \sum_{m=1}^M p_m\mathbb{E}\left[\left\lVert \sum_{i=0}^{R_r\tau-1} \nabla F_m(\boldsymbol{\theta}_{m,r}^{i})-\nabla F_m(\boldsymbol{\theta}_{r}^{0})\right\rVert^2\right]\\
      &\hspace{5mm}+2\eta^2 \sum_{m=1}^M p_m \mathbb{E}\left[\left\lVert \sum_{i=0}^{R_r\tau-1}\nabla F_m(\boldsymbol{\theta}_{r}^{0}) \right\rVert^2\right]\\
      &\overset{\text{A.1}}\leq \sum_{m=1}^M p_m^2 J_{m,r}^2+\eta^2 R_r\tau \sigma^2 \sum_{m=1}^M p_m^2
      +2\eta^2 R_r\tau\sum_{m=1}^M p_m \sum_{i=0}^{R_r\tau-1} L^2\mathbb{E}\left[\left\lVert\boldsymbol{\theta}_{m,r}^{i}-\boldsymbol{\theta}_{r}^{0}\right\rVert^2\right]\\
      &\hspace{5mm}+2\eta^2 R_r^2\tau^2\sum_{m=1}^M p_m \mathbb{E}\left[\left\lVert\nabla F_m(\boldsymbol{\theta}_{r}^{0})-F(\boldsymbol{\theta}_{r}^{0})+F(\boldsymbol{\theta}_{r}^{0}) \right\rVert^2\right]\\
      &\overset{\text{(i)}}\leq \sum_{m=1}^M p_m^2 J_{m,r}^2+\eta^2 R_r\tau \sigma^2 \sum_{m=1}^M p_m^2
      +2\eta^2L^2 R_r\tau\sum_{m=1}^M p_m \sum_{i=0}^{R_r\tau-1} \mathbb{E}\left[\left\lVert\boldsymbol{\theta}_{m,r}^{i}-\boldsymbol{\theta}_{r}^{0}\right\rVert^2\right]\\
      &\hspace{5mm}+4\eta^2 R_r^2\tau^2\sum_{m=1}^M p_m \mathbb{E}\left[\left\lVert\nabla F_m(\boldsymbol{\theta}_{r}^{0})-F(\boldsymbol{\theta}_{r}^{0}) \right\rVert^2\right]
      +4\eta^2 R_r^2\tau^2\sum_{m=1}^M p_m \mathbb{E}\left[\left\lVert\nabla F(\boldsymbol{\theta}_{r}^{0}) \right\rVert^2\right]\\
      &\overset{\text{(j)}}\leq \eta^2\sum_{m=1}^M p_m^2 J_{m,r}^2+\eta^2 R_r\tau \sigma^2 \sum_{m=1}^M p_m^2
      +2\eta^2 R_r\tau\sum_{m=1}^M p_m \sum_{i=0}^{R_r\tau-1} L^2\mathbb{E}\left[\left\lVert\boldsymbol{\theta}_{m,r}^{i}-\boldsymbol{\theta}_{r}^{0}\right\rVert^2\right]\\
      &\hspace{3mm}+4\eta^2 R_r^2\tau^2\hspace{-1mm}\sum_{m=1}^M p_m D_m^2+4\eta^2 R_r^2\tau^2 \mathbb{E}\left[\left\lVert\nabla F(\boldsymbol{\theta}_{r}^{0}) \right\rVert^2\right],
      \numberthis
  \end{align*}
  where the equality $\text{(d)}, \text{(f)}$ follows the fact that $\mathbb{E}[\lVert x \rVert^2]=\lVert \mathbb{E}[x]\rVert^2+\mathbb{E}[\lVert x-\mathbb{E}[x] \rVert^2]$, the equality $\text{(e)}, \text{(g)}$ is due to unbiased estimation and lemma 2 in \cite{wang2020tackling}, $\text{(h)}, \text{(i)}$ follows the fact that $ \lVert a+b \rVert^2 \leq 2\lVert a \rVert^2 +2\lVert b \rVert^2 $, $\text{(j)}$ follows A.3 and L.1.
  \section{proof of lemma 4} \label{Appendix:Lemma4}
\begin{proof}[Proof 1]
    \begin{align*}
    &\mathbb{E}\left[\left\lVert \boldsymbol{\theta}_{m,r}^0- \boldsymbol{\theta}_{m,r}^{i} \right\rVert^2\right]=\eta^2 \mathbb{E}\left[\left\lVert \sum_{s=0}^{i-1}\nabla F_m(\boldsymbol{\theta}_{m,r}^{s},\xi_{m,r}^{s}) \right\rVert^2\right]\\
    &\overset{\text{(k)}}= \eta^2 \mathbb{E}\left[\left\lVert \sum_{s=0}^{i-1} \left( \nabla F_m(\boldsymbol{\theta}_{m,r}^{s},\xi_{m,r}^{s}) -\nabla F_m(\boldsymbol{\theta}_{m,r}^{s})\right) \right\rVert^2\right]
    + \eta^2\mathbb{E}\left[\left\lVert \sum_{s=0}^{i-1} \nabla F_m(\boldsymbol{\theta}_{m,r}^{s}) \right\rVert^2\right]\\
    &\overset{\text{(l)}}=\eta^2\sum_{s=0}^{i-1} \mathbb{E}\left[\left\lVert  \nabla F_m(\boldsymbol{\theta}_{m,r}^{s},\xi_{m,r}^{s}) -\nabla F_m(\boldsymbol{\theta}_{m,r}^{s})\right\rVert^2\right]
    + \eta^2\mathbb{E}\left[\left\lVert \sum_{s=0}^{i-1}\nabla F_m(\boldsymbol{\theta}_{m,r}^{s}) \right\rVert^2\right]\\
    &\overset{\text{(m)}}\leq \eta^2 i \sigma^2 + \eta^2 i \sum_{s=0}^{i-1} \mathbb{E}\left[\left\lVert \nabla F_m(\boldsymbol{\theta}_{m,r}^{s}) \right\rVert^2\right]\\
    &=\eta^2 i \sigma^2 + \eta^2 i\sum_{s=0}^{i-1} \mathbb{E}\left[\left\lVert \nabla F_m(\boldsymbol{\theta}_{m,r}^{s}) - \nabla F_m(\boldsymbol{\theta}_{m,r}^{0})+\nabla F_m(\boldsymbol{\theta}_{m,r}^{0})\right\rVert^2\right]\\
    &\overset{\text{(n)}}\leq \eta^2 i \sigma^2 + 2\eta^2 i \sum_{s=0}^{i-1} \mathbb{E}\left[\left\lVert \nabla F_m(\boldsymbol{\theta}_{m,r}^{s}) - \nabla F_m(\boldsymbol{\theta}_{m,r}^{0})\right\rVert^2\right]\\
    &\hspace{5mm}+2\eta^2 i \sum_{s=0}^{i-1} \mathbb{E}\left[\left\lVert \nabla F_m(\boldsymbol{\theta}_{m,r}^{0})-\nabla F(\boldsymbol{\theta}_{r}^{0})+\nabla F(\boldsymbol{\theta}_{r}^{0})\right\rVert^2\right]\\
    &\overset{\text{A.1}}\leq  \eta^2 i \sigma^2 + 2\eta^2 i L^2 \sum_{s=0}^{i-1} \mathbb{E}\left[\left\lVert \boldsymbol{\theta}_{m,r}^{s} - \boldsymbol{\theta}_{m,r}^{0}\right\rVert^2\right]
    + 4 \eta^2 i  \sum_{s=0}^{i-1} \mathbb{E}\left[\left\lVert \nabla F_m(\boldsymbol{\theta}_{m,r}^{0})-\nabla F(\boldsymbol{\theta}_{r}^{0})\right\rVert^2\right]\\
    &\hspace{5mm}+ 4 \eta^2 i \sum_{s=0}^{i-1} \mathbb{E}\left[\left\lVert \nabla F(\boldsymbol{\theta}_{r}^{0})\right\rVert^2\right]\\
    &\overset{\text{(o)}}\leq  \eta^2 i \sigma^2 + 2\eta^2 i L^2 \sum_{s=0}^{i-1} \mathbb{E}\left[\left\lVert \boldsymbol{\theta}_{m,r}^{s} - \boldsymbol{\theta}_{m,r}^{0}\right\rVert^2\right]
    + 4 \eta^2 i^2 D_m^2  + 4 \eta^2 i^2 \mathbb{E}\left[\left\lVert \nabla F(\boldsymbol{\theta}_{r}^{0})\right\rVert^2\right],
    \numberthis
\end{align*}
where the equality $\text{(k)}$ follows $\mathbb{E}[\lVert x \rVert^2]=\lVert \mathbb{E}[x]\rVert^2+\mathbb{E}[\lVert x-\mathbb{E}[x] \rVert^2]$, $\text{(l)}$ is due to unbiased estimation and Lemma 2 in \cite{wang2020tackling}, $\text{(m)}$ follows A.2, the convexity of $l_2$-norm and Jensen's inequality, $\text{(n)}$ follows $ \lVert a+b \rVert^2 \leq 2\lVert a \rVert^2 +2\lVert b \rVert^2 $, $\text{(o)}$ ultilizes the fact $\boldsymbol{\theta}_{m,r}^{0}=\boldsymbol{\theta}_{r}^{0}$ and A.3.

\end{proof}

\begin{proof}[Proof 2]
Next, we prove the bound for the sum:  $\sum_{i=0}^{R_r\tau-1}  \mathbb{E}\left[\left\lVert \boldsymbol{\theta}_{m,r}^0- \boldsymbol{\theta}_{m,r}^{i} \right\rVert^2\right]$.
\begin{align*}
    & \sum_{i=0}^{R_r\tau-1}  \mathbb{E}\left[\left\lVert \boldsymbol{\theta}_{m,r}^0- \boldsymbol{\theta}_{m,r}^{i} \right\rVert^2\right]\\
    & \leq \sum_{i=0}^{R_r\tau-1} \Bigg(  \eta^2 i \sigma^2 + 2\eta^2 i L^2 \sum_{s=0}^{i-1} \mathbb{E}\left[\left\lVert \boldsymbol{\theta}_{m,r}^{s} - \boldsymbol{\theta}_{m,r}^{0}\right\rVert^2\right]
    + 4 \eta^2 i^2 D_m^2  + 4 \eta^2 i^2 \mathbb{E}\left[\left\lVert \nabla F(\boldsymbol{\theta}_{r}^{0})\right\rVert^2\right] \Bigg)\\
    &\overset{\text{(p)}}\leq  \frac{1}{2}R_r\tau(R_r\tau+1) \eta^2 \sigma^2
    +\eta^2 L^2 R_r\tau(R_r\tau+1) \sum_{s=0}^{R_r\tau-1} \mathbb{E}\left[\left\lVert \boldsymbol{\theta}_{m,r}^{s} - \boldsymbol{\theta}_{m,r}^{0}\right\rVert^2\right]\\
    &\hspace{5mm}+\frac{2}{3}R_r\tau(R_r\tau+1)(2R_r\tau+1) \eta^2 D_m^2+\frac{2}{3}R_r\tau(R_r\tau+1)(2R_r\tau+1) \eta^2 \mathbb{E}\left[\left\lVert \nabla F(\boldsymbol{\theta}_{r}^{0})\right\rVert^2\right],
    \numberthis
    \label{proof:lemma3}
\end{align*}
where the equality $\text{(p)}$ follows that $\sum_{i=1}^{n}i=\frac{n(n+1)}{2}$, $\sum_{i=1}^{n}i^2=\frac{n(n+1)(2n+1)}{6}$, and that $i\leq R_r\tau-1$. Both sides of Eq.\ref{proof:lemma3} contain $\sum_{i=0}^{R_r\tau-1}\mathbb{E}\left[\left\lVert \boldsymbol{\theta}_{m,r}^0- \boldsymbol{\theta}_{m,r}^{i} \right\rVert^2\right]$, by minor arrangement,
\begin{align*}
    & (1-R_r\tau(R_r\tau+1)\eta^2L^2)\sum_{i=0}^{R_r\tau-1} \mathbb{E}\left[\left\lVert \boldsymbol{\theta}_{m,r}^0- \boldsymbol{\theta}_{m,r}^{i} \right\rVert^2\right]\\
    &\leq\frac{1}{2}R_r\tau(R_r\tau+1) \eta^2 \sigma^2+\frac{2}{3}R_r\tau(R_r\tau+1)(2R_r\tau+1) \eta^2 D_m^2\\
    &\hspace{5mm}+\frac{2}{3}R_r\tau(R_r\tau+1)(2R_r\tau+1) \eta^2 \mathbb{E}\left[\left\lVert \nabla F(\boldsymbol{\theta}_{r}^{0})\right\rVert^2\right].
    \numberthis
\end{align*}
With $1-R_r\tau(R_r\tau+1)\eta^2L^2>0$, we reach the second result in L.4.

\end{proof}

\end{document}

%% file: outage.tex
\begin{tikzpicture}[scale=1\columnwidth/10cm,font=\footnotesize]
\begin{axis}[%
width=6cm,
height=4cm,
scale only axis,
xmin=1,
xmax=15,
ymode=log,
ymin=0,
ymax=1,
xlabel={SNR},
ylabel={Probability},
y label style={at={(axis description cs:0.09,0.75)},
                     rotate=0, anchor=south east},
x label style={at={(axis description cs:0.55,-0.07)},
                     rotate=0, anchor=south east},
yminorticks=true,
xminorticks=true,
axis background/.style={fill=white},
xmajorgrids,
xminorgrids,
ymajorgrids,
yminorgrids,
legend style={at={(-0.3,1.1)}, anchor=south west, legend columns=3, legend cell align=left, align=left, draw=white!15!black}
]


\addplot [color=fuchsiapink, line width=1.5pt, mark size=2pt, mark=square*, mark options={solid, fill=fuchsiapink, draw=fuchsiapink}]
  table[row sep=crcr]{%
1	0.319507910772894\\
2	0.159753955386447\\
3	0.106502636924298\\
4	0.0798769776932236\\
5	0.0639015821545788\\
6	0.0532513184621490\\
7	0.0456439872532706\\
8	0.0399384888466118\\
9	0.0355008789747660\\
10	0.0319507910772894\\
11	0.0290461737066267\\
12	0.0266256592310745\\
13	0.0245775315979149\\
14	0.0228219936266353\\
15	0.0213005273848596\\
};
\addlegendentry{$q(R=0.2)$}

\addplot [color=fuchsiapink, line width=1.5pt, mark size=2pt, mark=square*, mark options={solid, fill=fuchsiapink, draw=fuchsiapink},dashed]
  table[row sep=crcr]{%
1	0.998416697976311\\
2	0.747277373411601\\
3	0.355460776675594\\
4	0.155027306122468\\
5	0.0703892441590062\\
6	0.0341262980069923\\
7	0.0176539118058114\\
8	0.00967792276657445\\
9	0.00557981314821277\\
10	0.00336027188586308\\
11	0.00210138324350312\\
12	0.00135795399827842\\
13	0.000903109067117171\\
14	0.000616010795778981\\
15	0.000429718474990770\\
};
\addlegendentry{$P_O(R=0.2,s=5)$}

\addplot[color=dollarbill, line width=1.5pt, mark size=1.5pt, mark=*, mark options={solid, fill=dollarbill, draw=dollarbill}]
  table[row sep=crcr]{%
1	0.148698354997035\\
2	0.0743491774985176\\
3	0.0495661183323450\\
4	0.0371745887492588\\
5	0.0297396709994070\\
6	0.0247830591661725\\
7	0.0212426221424336\\
8	0.0185872943746294\\
9	0.0165220394441150\\
10	0.0148698354997035\\
11	0.0135180322724577\\
12	0.0123915295830863\\
13	0.0114383349997719\\
14	0.0106213110712168\\
15	0.00991322366646901\\
};
\addlegendentry{$q(R=0.1)$}

\addplot [color=dollarbill, line width=1.5pt, mark size=1.5pt, mark=*, mark options={solid, fill=dollarbill, draw=dollarbill},dashed]
  table[row sep=crcr]{%
1	0.680800087264951\\
2	0.121818653134222\\
3	0.0252389354739222\\
4	0.00693572202026379\\
5	0.00236244860622312\\
6	0.000942508588244475\\
7	0.000423623935967844\\
8	0.000208858535706201\\
9	0.000110843854733053\\
10	6.24597168406849e-05\\
11	3.69880693811916e-05\\
12	2.28393730904998e-05\\
13	1.46150244519406e-05\\
14	9.64459445086354e-06\\
15	6.53762788940468e-06\\
};
\addlegendentry{$P_O(R=0.1, s=5)$}

\addplot[color=bleudefrance, line width=1.5pt, mark size=1.5pt, mark=triangle*, mark options={solid, fill=dollarbill, draw=bleudefrance}, dashed]
  table[row sep=crcr]
 {
1	0.990858235565220\\
2	0.530230725863419\\
3	0.157489366753013\\
4	0.0456536386567069\\
5	0.0146667885682311\\
6	0.00528165483384368\\
7	0.00210673220911108\\
8	0.000917015809947819\\
9	0.000429788815632337\\
10	0.000214488836887485\\
11	0.000112948697503138\\
12	6.22983031346411e-05\\
13	3.57739181670307e-05\\
14	2.12809408374845e-05\\
15	1.30602970225607e-05\\
};
\addlegendentry{$P_O(R=0.2, s=7)$}

\end{axis}
\end{tikzpicture}

%% file: test_acc_iid.tex
\begin{tikzpicture}[scale=1\columnwidth/10cm,font=\footnotesize]
\begin{axis}[%
width=6cm,
height=4cm,
scale only axis,
xmin=1,
xmax=20,
ymin=0.1,
ymax=1,
xlabel={Communication Round},
ylabel={Test Accuracy},
y label style={at={(axis description cs:0.12,0.9)},
                     rotate=0, anchor=south east},
x label style={at={(axis description cs:0.7,-0.07)},
                     rotate=0, anchor=south east},      
yminorticks=true,
xminorticks=true,
axis background/.style={fill=white},
xmajorgrids,
xmajorgrids,
ymajorgrids,
yminorgrids,
legend style={at={(-0.3,1.05)}, anchor=south west, legend columns=3, legend cell align=left, align=left, draw=white!15!black}
]


\addplot [color=mikadoyellow, line width=1.5pt, mark size=2pt, mark=square*, mark options={solid, fill=mikadoyellow, draw=mikadoyellow}]
  table[row sep=crcr]{%
    1.0000    0.2141\\
    2.0000    0.3944\\
    3.0000    0.5439\\
    4.0000    0.6575\\
    5.0000    0.7220\\
    6.0000    0.7741\\
    7.0000    0.8145\\
    8.0000    0.8424\\
    9.0000    0.8606\\
   10.0000    0.8747\\
   11.0000    0.8846\\
   12.0000    0.8921\\
   13.0000    0.8986\\
   14.0000    0.9039\\
   15.0000    0.9084\\
   16.0000    0.9121\\
   17.0000    0.9153\\
   18.0000    0.9182\\
   19.0000    0.9209\\
   20.0000    0.9234\\
};
\addlegendentry{QFL}

\addplot [color=fuchsiapink, line width=1.5pt, mark=diamond*]
  table[row sep=crcr]{%
1	0.249840000000000\\
2	0.399110000000000\\
3	0.515480000000000\\
4	0.629360000000000\\
5	0.717540000000000\\
6	0.778260000000000\\
7	0.814810000000000\\
8	0.838190000000000\\
9	0.854110000000000\\
10	0.866160000000000\\
11	0.875640000000000\\
12	0.883140000000000\\
13	0.889400000000000\\
14	0.894880000000000\\
15	0.899630000000000\\
16	0.903930000000000\\
17	0.907520000000000\\
18	0.910680000000000\\
19	0.913730000000000\\
20	0.916000000000000\\
};
\addlegendentry{Non-blind FL, SNR=5}

\addplot [color=azure(colorwheel), line width=1.5pt, mark size=1.5pt, mark=triangle*, mark options={solid, fill=azure(colorwheel), draw=azure(colorwheel)},dashed]
  table[row sep=crcr]{%
    1.0000    0.2280\\
    2.0000    0.3814\\
    3.0000    0.5325\\
    4.0000    0.6148\\
    5.0000    0.6766\\
    6.0000    0.7435\\
    7.0000    0.7927\\
    8.0000    0.8307\\
    9.0000    0.8545\\
   10.0000    0.8692\\
   11.0000    0.8806\\
   12.0000    0.8862\\
   13.0000    0.8879\\
   14.0000    0.8977\\
   15.0000    0.9004\\
   16.0000    0.9080\\
   17.0000    0.9170\\
   18.0000    0.9148\\
   19.0000    0.9185\\
   20.0000    0.9212\\
};
\addlegendentry{CoGC, SNR=3}

\addplot [color=dollarbill, line width=1.5pt, mark size=1.5pt, mark=*, mark options={solid, fill=dollarbill, draw=dollarbill},dashed]
  table[row sep=crcr]{%
    1.0000    0.1783\\
    2.0000    0.2906\\
    3.0000    0.3645\\
    4.0000    0.4368\\
    5.0000    0.5088\\
    6.0000    0.5672\\
    7.0000    0.6111\\
    8.0000    0.6739\\
    9.0000    0.7054\\
   10.0000    0.7407\\
   11.0000    0.7716\\
   12.0000    0.7917\\
   13.0000    0.8149\\
   14.0000    0.8271\\
   15.0000    0.8372\\
   16.0000    0.8438\\
   17.0000    0.8489\\
   18.0000    0.8513\\
   19.0000    0.8550\\
   20.0000    0.8572\\
};
\addlegendentry{Blind FL, SNR=5}

\addplot [color=dollarbill, line width=1.5pt, mark size=1.5pt, mark=*, mark options={solid, fill=dollarbill, draw=dollarbill}]
  table[row sep=crcr]{%
1.0000    0.2085\\
    2.0000    0.2823\\
    3.0000    0.3504\\
    4.0000    0.4219\\
    5.0000    0.4541\\
    6.0000    0.4882\\
    7.0000    0.5204\\
    8.0000    0.5509\\
    9.0000    0.5916\\
   10.0000    0.6268\\
   11.0000    0.6538\\
   12.0000    0.6670\\
   13.0000    0.6724\\
   14.0000    0.6797\\
   15.0000    0.6820\\
   16.0000    0.6993\\
   17.0000    0.7020\\
   18.0000    0.7156\\
   19.0000    0.7303\\
   20.0000    0.7438\\
};
\addlegendentry{Blind FL, SNR=3}

\addplot[color=fuchsiapink, line width=1.5pt, mark size=1.5pt, mark=diamond*, mark options={solid, fill=mikadoyellow, draw=fuchsiapink}, dashed]
  table[row sep=crcr]
 {
1	0.228700000000000\\
2	0.405810000000000\\
3	0.553350000000000\\
4	0.649460000000000\\
5	0.720370000000000\\
6	0.771450000000000\\
7	0.810280000000000\\
8	0.836730000000000\\
9	0.852810000000000\\
10	0.865470000000000\\
11	0.875360000000000\\
12	0.883550000000000\\
13	0.890220000000000\\
14	0.896060000000000\\
15	0.900480000000000\\
16	0.904500000000000\\
17	0.908410000000000\\
18	0.912220000000000\\
19	0.915420000000000\\
20	0.918040000000000\\
};
\addlegendentry{Non-blind FL, SNR=3}

\addplot [color=azure(colorwheel), line width=1.5pt, mark size=1.5pt, mark=triangle*, mark options={solid, fill=azure(colorwheel), draw=azure(colorwheel)}]
  table[row sep=crcr]{%
    1.0000    0.1782\\
    2.0000    0.3356\\
    3.0000    0.5050\\
    4.0000    0.5950\\
    5.0000    0.6972\\
    6.0000    0.7624\\
    7.0000    0.7910\\
    8.0000    0.8263\\
    9.0000    0.8370\\
   10.0000    0.8553\\
   11.0000    0.8654\\
   12.0000    0.8780\\
   13.0000    0.8881\\
   14.0000    0.8936\\
   15.0000    0.8998\\
   16.0000    0.9043\\
   17.0000    0.9083\\
   18.0000    0.9122\\
   19.0000    0.9139\\
   20.0000    0.9176\\
};
\addlegendentry{CoGC, SNR=5}

\end{axis}
\end{tikzpicture}

%% file: test_acc_noniid_new.tex
\begin{tikzpicture}[scale=1\columnwidth/10cm,font=\footnotesize]
\begin{axis}[%
width=6cm,
height=4cm,
scale only axis,
xmin=1,
xmax=20,
ymin=0.1,
ymax=0.9,
xlabel={Communication Round},
ylabel={Test Accuracy},
y label style={at={(axis description cs:0.12,0.9)},
                     rotate=0, anchor=south east},
x label style={at={(axis description cs:0.7,-0.07)},
                     rotate=0, anchor=south east},      
yminorticks=true,
xminorticks=true,
axis background/.style={fill=white},
xmajorgrids,
xmajorgrids,
ymajorgrids,
yminorgrids,
legend style={at={(-0.3,1.05)}, anchor=south west, legend columns=3, legend cell align=left, align=left, draw=white!15!black}
]


%
\addplot [color=mikadoyellow, line width=1.5pt, mark size=1.5pt, mark=square*, mark options={solid, fill=mikadoyellow, draw=mikadoyellow},dashed]
  table[row sep=crcr]{%
1	0.208400000000000\\
2	0.363900000000000\\
3	0.472400000000000\\
4	0.548800000000000\\
5	0.604700000000000\\
6	0.633200000000000\\
7	0.656700000000000\\
8	0.680900000000000\\
9	0.700700000000000\\
10	0.729900000000000\\
11	0.741300000000000\\
12	0.757500000000000\\
13	0.765800000000000\\
14	0.773000000000000\\
15	0.780700000000000\\
16	0.788600000000000\\
17	0.796100000000000\\
18	0.797600000000000\\
19	0.811900000000000\\
20	0.814300000000000\\
};
\addlegendentry{QFL, 5 class}

\addplot [color=mikadoyellow, line width=1.5pt, mark=square*]
  table[row sep=crcr]{%
1	0.111400000000000\\
2	0.223700000000000\\
3	0.303700000000000\\
4	0.356550000000000\\
5	0.397650000000000\\
6	0.449850000000000\\
7	0.478450000000000\\
8	0.500100000000000\\
9	0.528700000000000\\
10	0.556150000000000\\
11	0.572950000000000\\
12	0.594250000000000\\
13	0.603650000000000\\
14	0.623900000000000\\
15	0.632250000000000\\
16	0.641400000000000\\
17	0.658350000000000\\
18	0.666650000000000\\
19	0.679050000000000\\
20	0.686800000000000\\
};
\addlegendentry{QFL, 1 classes}


%

\addplot [color=dollarbill, line width=1.5pt, mark size=1.5pt, mark=*, mark options={solid, fill=dollarbill, draw=dollarbill}]
  table[row sep=crcr]{%
1	0.128468750000000\\
2	0.153437500000000\\
3	0.214212500000000\\
4	0.200956250000000\\
5	0.223187500000000\\
6	0.245581250000000\\
7	0.232312500000000\\
8	0.236393750000000\\
9	0.215381250000000\\
10	0.234781250000000\\
11	0.228343750000000\\
12	0.181056250000000\\
13	0.180200000000000\\
14	0.212306250000000\\
15	0.207037500000000\\
16	0.184281250000000\\
17	0.164706250000000\\
18	0.181950000000000\\
19	0.191150000000000\\
20	0.172106250000000\\
};
\addlegendentry{Blind FL, 1 class}

%

%

\addplot [color=fuchsiapink, line width=1.5pt, mark size=1.5pt, mark=diamond*, mark options={solid, fill=fuchsiapink, draw=fuchsiapink},dashed]
  table[row sep=crcr]{%
    1.0000    0.1879\\
    2.0000    0.2994\\
    3.0000    0.3677\\
    4.0000    0.4454\\
    5.0000    0.5221\\
    6.0000    0.5658\\
    7.0000    0.6142\\
    8.0000    0.6475\\
    9.0000    0.6573\\
   10.0000    0.6914\\
   11.0000    0.7090\\
   12.0000    0.7151\\
   13.0000    0.7210\\
   14.0000    0.7451\\
   15.0000    0.7590\\
   16.0000    0.7673\\
   17.0000    0.7508\\
   18.0000    0.7816\\
   19.0000    0.7852\\
   20.0000    0.7777\\
};
\addlegendentry{Non-blind FL, 5 classes}

\addplot [color=dollarbill, line width=1.5pt, mark size=1.5pt, mark=*, mark options={solid, fill=dollarbill, draw=dollarbill},dashed]
  table[row sep=crcr]{%
    1.0000    0.2048\\
    2.0000    0.3033\\
    3.0000    0.3956\\
    4.0000    0.4209\\
    5.0000    0.4585\\
    6.0000    0.4632\\
    7.0000    0.5069\\
    8.0000    0.5225\\
    9.0000    0.5168\\
   10.0000    0.5367\\
   11.0000    0.5292\\
   12.0000    0.5418\\
   13.0000    0.5264\\
   14.0000    0.5393\\
   15.0000    0.5555\\
   16.0000    0.5687\\
   17.0000    0.5582\\
   18.0000    0.5478\\
   19.0000    0.5625\\
   20.0000    0.5560\\
};
\addlegendentry{Blind FL, 5 classes}

\addplot[color=azure(colorwheel), line width=1.5pt, mark size=1.5pt, mark=triangle*, mark options={solid, fill=azure(colorwheel), draw=azure(colorwheel)}, dashed]
  table[row sep=crcr]{%
    1.0000    0.2124\\
    2.0000    0.3475\\
    3.0000    0.4658\\
    4.0000    0.5392\\
    5.0000    0.5900\\
    6.0000    0.6329\\
    7.0000    0.6667\\
    8.0000    0.6907\\
    9.0000    0.7126\\
   10.0000    0.7285\\
   11.0000    0.7439\\
   12.0000    0.7539\\
   13.0000    0.7617\\
   14.0000    0.7716\\
   15.0000    0.7755\\
   16.0000    0.7838\\
   17.0000    0.7894\\
   18.0000    0.7967\\
   19.0000    0.8027\\
   20.0000    0.8071\\
};
\addlegendentry{CoGC, 5 classes}

\addplot [color=fuchsiapink, line width=1.5pt, mark=diamond*]
  table[row sep=crcr]
 {
    1.0000    0.1206\\
    2.0000    0.1767\\
    3.0000    0.2126\\
    4.0000    0.2799\\
    5.0000    0.2914\\
    6.0000    0.3412\\
    7.0000    0.3641\\
    8.0000    0.3951\\
    9.0000    0.4156\\
   10.0000    0.4361\\
   11.0000    0.4574\\
   12.0000    0.4857\\
   13.0000    0.5072\\
   14.0000    0.5173\\
   15.0000    0.5489\\
   16.0000    0.5612\\
   17.0000    0.5762\\
   18.0000    0.5846\\
   19.0000    0.5910\\
   20.0000    0.6048\\
};
\addlegendentry{Non-blind FL, 1 class}

\addplot [color=azure(colorwheel), line width=1.5pt, mark=triangle*]
  table[row sep=crcr]{%
    1.0000    0.1386\\
    2.0000    0.2030\\
    3.0000    0.2892\\
    4.0000    0.3505\\
    5.0000    0.3949\\
    6.0000    0.4430\\
    7.0000    0.4750\\
    8.0000    0.4968\\
    9.0000    0.5230\\
   10.0000    0.5417\\
   11.0000    0.5657\\
   12.0000    0.5785\\
   13.0000    0.6023\\
   14.0000    0.6143\\
   15.0000    0.6290\\
   16.0000    0.6455\\
   17.0000    0.6497\\
   18.0000    0.6657\\
   19.0000    0.6648\\
   20.0000    0.6784\\
};
\addlegendentry{CoGC, 1 class}

\end{axis}
\end{tikzpicture}

%% file: reference.bbl

%% file: main_arxiv.bbl
\begin{thebibliography}{10}
\providecommand{\url}[1]{#1}
\csname url@samestyle\endcsname
\providecommand{\newblock}{\relax}
\providecommand{\bibinfo}[2]{#2}
\providecommand{\BIBentrySTDinterwordspacing}{\spaceskip=0pt\relax}
\providecommand{\BIBentryALTinterwordstretchfactor}{4}
\providecommand{\BIBentryALTinterwordspacing}{\spaceskip=\fontdimen2\font plus
\BIBentryALTinterwordstretchfactor\fontdimen3\font minus \fontdimen4\font\relax}
\providecommand{\BIBforeignlanguage}[2]{{%
\expandafter\ifx\csname l@#1\endcsname\relax
\typeout{** WARNING: IEEEtran.bst: No hyphenation pattern has been}%
\typeout{** loaded for the language `#1'. Using the pattern for}%
\typeout{** the default language instead.}%
\else
\language=\csname l@#1\endcsname
\fi
#2}}
\providecommand{\BIBdecl}{\relax}
\BIBdecl

\bibitem{niknam2020federated}
S.~Niknam, H.~S. Dhillon, and J.~H. Reed, ``Federated learning for wireless communications: Motivation, opportunities, and challenges,'' \emph{IEEE Communications Magazine}, vol.~58, no.~6, pp. 46--51, 2020.

\bibitem{zhang2021survey}
C.~Zhang, Y.~Xie, H.~Bai, B.~Yu, W.~Li, and Y.~Gao, ``A survey on federated learning,'' \emph{Knowledge-Based Systems}, vol. 216, p. 106775, 2021.

\bibitem{kairouz2021advances}
P.~Kairouz, H.~B. McMahan, B.~Avent, A.~Bellet, M.~Bennis, A.~N. Bhagoji, K.~Bonawitz, Z.~Charles, G.~Cormode, R.~Cummings \emph{et~al.}, ``Advances and open problems in federated learning,'' \emph{Foundations and Trends{\textregistered} in Machine Learning}, vol.~14, no. 1--2, pp. 1--210, 2021.

\bibitem{9261995}
C.~T. Dinh, N.~H. Tran, M.~N.~H. Nguyen, C.~S. Hong, W.~Bao, A.~Y. Zomaya, and V.~Gramoli, ``Federated learning over wireless networks: Convergence analysis and resource allocation,'' \emph{IEEE/ACM Transactions on Networking}, vol.~29, no.~1, pp. 398--409, 2021.

\bibitem{xhemrishi2022computational}
M.~Xhemrishi, A.~G. i~Amat, E.~Rosnes, and A.~Wachter-Zeh, ``Computational code-based privacy in coded federated learning,'' in \emph{2022 IEEE International Symposium on Information Theory (ISIT)}.\hskip 1em plus 0.5em minus 0.4em\relax IEEE, 2022, pp. 2034--2039.

\bibitem{amiri2020federated}
M.~M. Amiri, D.~Gunduz, S.~R. Kulkarni, and H.~V. Poor, ``Federated learning with quantized global model updates,'' 2020.

\bibitem{9515709}
M.~M. Amiri, D.~Gündüz, S.~R. Kulkarni, and H.~V. Poor, ``Convergence of federated learning over a noisy downlink,'' \emph{IEEE Transactions on Wireless Communications}, vol.~21, no.~3, pp. 1422--1437, 2022.

\bibitem{9563232}
H.~Xing, O.~Simeone, and S.~Bi, ``Federated learning over wireless device-to-device networks: Algorithms and convergence analysis,'' \emph{IEEE Journal on Selected Areas in Communications}, vol.~39, no.~12, pp. 3723--3741, 2021.

\bibitem{wang2021quantized}
Y.~Wang, Y.~Xu, Q.~Shi, and T.-H. Chang, ``Quantized federated learning under transmission delay and outage constraints,'' \emph{IEEE Journal on Selected Areas in Communications}, vol.~40, no.~1, pp. 323--341, 2021.

\bibitem{saha2022colrel}
R.~Saha, M.~Yemini, E.~Ozfatura, D.~Gunduz, and A.~Goldsmith, ``Colrel: Collaborative relaying for federated learning over intermittently connected networks,'' in \emph{Workshop on Federated Learning: Recent Advances and New Challenges (in Conjunction with NeurIPS 2022)}, 2022.

\bibitem{9311931}
M.~Chen, H.~V. Poor, W.~Saad, and S.~Cui, ``Wireless communications for collaborative federated learning,'' \emph{IEEE Communications Magazine}, vol.~58, no.~12, pp. 48--54, 2020.

\bibitem{schlegel2023codedpaddedfl}
R.~Schlegel, S.~Kumar, E.~Rosnes, and A.~G. i~Amat, ``Codedpaddedfl and codedsecagg: Straggler mitigation and secure aggregation in federated learning,'' \emph{IEEE Transactions on Communications}, 2023.

\bibitem{mcmahan2017communication}
B.~McMahan, E.~Moore, D.~Ramage, S.~Hampson, and B.~A. y~Arcas, ``Communication-efficient learning of deep networks from decentralized data,'' in \emph{Artificial intelligence and statistics}.\hskip 1em plus 0.5em minus 0.4em\relax PMLR, 2017, pp. 1273--1282.

\bibitem{5595117}
M.~Xiao and M.~Skoglund, ``Multiple-user cooperative communications based on linear network coding,'' \emph{IEEE Transactions on Communications}, vol.~58, no.~12, pp. 3345--3351, 2010.

\bibitem{tandon2017gradient}
R.~Tandon, Q.~Lei, A.~G. Dimakis, and N.~Karampatziakis, ``Gradient coding: Avoiding stragglers in distributed learning,'' in \emph{International Conference on Machine Learning}.\hskip 1em plus 0.5em minus 0.4em\relax PMLR, 2017, pp. 3368--3376.

\bibitem{lian2017can}
X.~Lian, C.~Zhang, H.~Zhang, C.-J. Hsieh, W.~Zhang, and J.~Liu, ``Can decentralized algorithms outperform centralized algorithms? a case study for decentralized parallel stochastic gradient descent,'' \emph{Advances in neural information processing systems}, vol.~30, 2017.

\bibitem{wang2020tackling}
J.~Wang, Q.~Liu, H.~Liang, G.~Joshi, and H.~V. Poor, ``Tackling the objective inconsistency problem in heterogeneous federated optimization,'' \emph{Advances in neural information processing systems}, vol.~33, pp. 7611--7623, 2020.

\bibitem{Wang_2022}
\BIBentryALTinterwordspacing
Y.~Wang, Y.~Xu, Q.~Shi, and T.-H. Chang, ``Quantized federated learning under transmission delay and outage constraints,'' \emph{IEEE Journal on Selected Areas in Communications}, vol.~40, no.~1, p. 323–341, Jan. 2022. [Online]. Available: \url{http://dx.doi.org/10.1109/JSAC.2021.3126081}
\BIBentrySTDinterwordspacing

\end{thebibliography}
